\documentclass{natureprintstyle}
\usepackage{graphicx,amsmath}
\usepackage{astjnlabbrev-nature}

\usepackage{amssymb}

\title{Formation of new stellar populations from gas accreted by massive young star clusters}

\author{Chengyuan Li$^{1,2,3}$, Richard de Grijs$^{1,4}$, Licai Deng$^{2}$, Aaron M. Geller$^{5,6}$, Yu Xin$^{2}$, Yi Hu$^{2}$ and Claude-Andr\'e Faucher-Gigu\'ere$^{5}$}
\begin{document}

\maketitle

\begin{affiliations}
 \item Kavli Institute for Astronomy \& Astrophysics, Peking
   University, Yi He Yuan Lu 5, Hai Dian District, Beijing 100871,
   China\\
 \item Key Laboratory for Optical Astronomy, National Astronomical
   Observatories, Chinese Academy of Sciences, 20A Datun Road,
   Chaoyang District, Beijing 100012, China\\
  \item Purple Mountain Observatory, Chines Academy of Sciences, 2 West Beijing Road, Nanjing, 210008, China\\
 \item International Space Science Institute-Beijing, 1 Nanertiao,
 Hai Dian District, Beijing 100190, China\\
 \item Center for Interdisciplinary Exploration and Research in
 Astrophysics (CIERA) and Department of Physics and Astronomy,
 Northwestern University, 2145 Sheridan Road, Evanston, IL 60208, USA\\
 \item Department of Astronomy and Astrophysics, University of Chicago,
 5640 S. Ellis Avenue, Chicago, IL 60637, USA\\
\end{affiliations}

\begin{abstract}
Stars in star clusters are thought to form in a single burst from a common progenitor cloud of molecular gas. However, massive, old ‘globular’ clusters -- with ages greater than 10 billion years and masses of several hundred thousand solar masses -- often harbour multiple stellar populations\cite{Grat12,Piot07,Milo12,Bedi04}, indicating that more than one star-forming event occurred during their lifetimes. Colliding stellar winds from late-stage, asymptotic-giant-branch stars\cite{DErc08,Bekk11,Renz08,Bast15} are often invoked as second-generation star-formation trigger. The initial cluster masses should be at least 10 times more massive than they are today for this to work\cite{Bast15}. However, large populations of clusters with masses greater than a few million solar masses are not found in the local Universe. Here we report on three 1-2 billion-year-old, massive star clusters in the Magellanic Clouds, which show clear evidence of burst-like star formation that occurred a few hundred million years after their initial formation era. We show that such clusters could accrete sufficient gas reservoirs to form new stars if the clusters orbited in their host galaxies’ gaseous discs throughout the period between their initial formation and the more recent bursts of star formation. This may eventually give rise to the ubiquitous multiple stellar populations in globular clusters.
\end{abstract}

The colour-magnitude diagrams of NGC 1783, NGC 1696 and NGC 411 are shown in Figure 1. These stellar distributions -- the observational counterparts of the theoretical Hertzsprung-Russell diagrams, which relate the stellar surface temperatures to their luminosities -- have been field-star decontaminated by careful application of statistical background-subtraction techniques (see Methods). Figure 1a shows that the majority of stars associated with NGC 1783 are well-described by an isochrone\cite{Mari08} -- a theoretical ridge-line that describes stars with identical ages but covering a range of initial masses -- characterized by an age of log($t$ yr$^{-1}$) = 9.15 (1.4 billion years; see Methods). However, one can also clearly discern two additional, bright stellar sequences, denoted ‘A’ and ‘B’, with younger ages of, respectively, log($t$ yr$^{-1}$) = 8.65 (450 million years) and log($t$ yr$^{-1}$) = 8.95 (890 million years). The latter appear to have similar chemical compositions as the cluster’s bulk stellar population (in particular in terms of the helium and heavier-element abundances), given the absence of any clear differences in the observed ridge-line colours. Sequence A may also include a subpopulation of stars associated with the cluster’s red clump, the 0.7 to 2 solar-mass analogues of the helium-burning horizontal-branch stars (indicated by the orange area). Figure 1b shows the colour-magnitude diagram pertaining to NGC 1696. It also exhibits a bright, young simple-stellar-population sequence characterized by log($t$ yr$^{-1}$) = 8.70 (500 million years), whereas the bulk of the NGC 1696 stars are best represented by an older age of log($t$ yr$^{-1}$) = 9.18 (1.5 billion years). Compared with the cluster’s dominant main sequence, the younger sequence exhibits a colour offset of approximately -0.06 mag (a shift to bluer colours), which is consistent with a stellar population characterized by an enhanced helium abundance of $Y$ = 0.330 (33.0\% of helium atoms by mass), while $Y$ = 0.256 for the bulk of the cluster stars. A similar result for NGC 411 is shown in Figure 1c, where we also find an additional, brighter stellar sequence that is well-represented by a younger isochrone of log(t yr-1) = 8.50 (320 million years) and $Y$ = 0.400, compared with log($t$ yr$^{-1}$) = 9.14 (1.4 billion years) and $Y$ = 0.252 for the bulk of the cluster’s stellar population. These well-populated, younger and helium-enhanced stellar sequences represent the strongest evidence yet that additional, post-formation starburst events may have occurred in our sample clusters. The enhanced helium abundances may also lead to small changes in the best-fitting cluster ages, but there are currently no appropriate model isochrones available to accurately explore this effect for stellar populations younger than 1 billion years. Nevertheless, the general sense of helium-enhanced young sequences shown here is robust. To date, no other massive clusters of equivalent age are known to host similarly significant populations of younger stars\cite{Bast13}.

\begin{figure*}
\hspace{-0.4cm}\centering
\includegraphics[width=2.0\columnwidth]{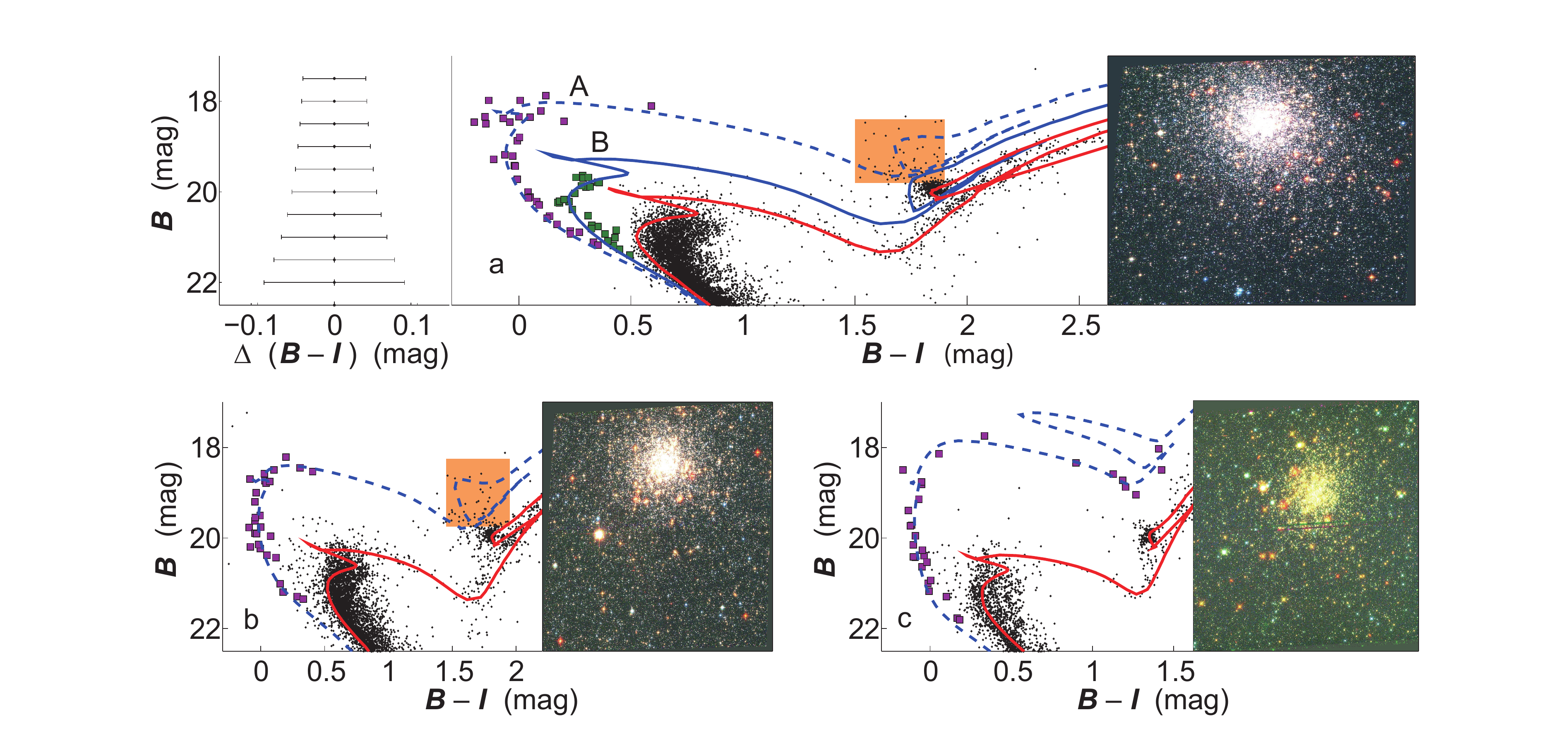}
\begin{center}
\caption{{\bf Colour--magnitude diagrams, including the best-fitting isochrones, and true-colour images for all three clusters.} a: NGC 1783. Purple and dark green squares: sequence-A and -B stars, respectively; red solid, blue solid and blue dashed lines: isochrones for  $\log (t \mbox{ yr}^{-1})$  = 9.15, 8.95 and 8.65, respectively. The orange box indicates the region where red-clump stars associated with sequence A may be found. Representative 1$\sigma$ measurement uncertainties are shown in the left-hand subpanel; note that its x-axis scale is different from that of the main panel. b: NGC 1696. Purple squares: younger sequence; red solid and blue dashed lines: isochrones for  $\log (t \mbox{ yr}^{-1})$  = 9.18 and 8.70, respectively. c: NGC 411. Purple squares: younger sequence; red solid and blue dashed lines: isochrones for  $\log (t \mbox{ yr}^{-1})$ = 9.14 and 8.50, respectively. The measurement uncertainties in panels b and c are equivalent to those shown for panel a. For the young sequences in NGC 1696 and NGC 411, enhanced helium abundances have been adopted (see text).}
\end{center}
\label{F1}
\end{figure*}

We explored whether `blue straggler stars' --  stars that have been rejuvenated through either stellar collisions or mass transfer in binary stellar systems\cite{Lu10,Hill76} -- could be entirely responsible for the presence of these younger sequences. If they are formed through mass transfer in unresolved, compact binary systems, they would appear brighter and slightly redder than the corresponding isochrone\cite{Li13} describing zero-age single stars (stars whose output luminosities are no longer powered by the excess energy gained from gravitational contraction but which are instead driven by nuclear fusion of hydrogen atoms). However, the younger sequences in both clusters are too blue to account for a binary origin and can instead be very well described by single-star isochrones. Given their young ages and the timescales involved in evolution through binary mass transfer, if any of these younger stars are indeed blue stragglers, they will most likely have formed through stellar collisions. Collisionally formed blue stragglers, like secondary stellar generations originating from colliding stellar winds, are expected to be more centrally concentrated than the clusters’ dominant (by number) stellar populations\cite{Bekk11}. Figure 2 compares the normalized radial distributions of the young sequences with those of ‘normal’ cluster stars of similar luminosity. The stars in the young sequences are markedly less centrally concentrated than the dominant older population of cluster members.

\begin{figure}
\hspace{-0.4cm}\centering
\includegraphics[width=1.0\columnwidth]{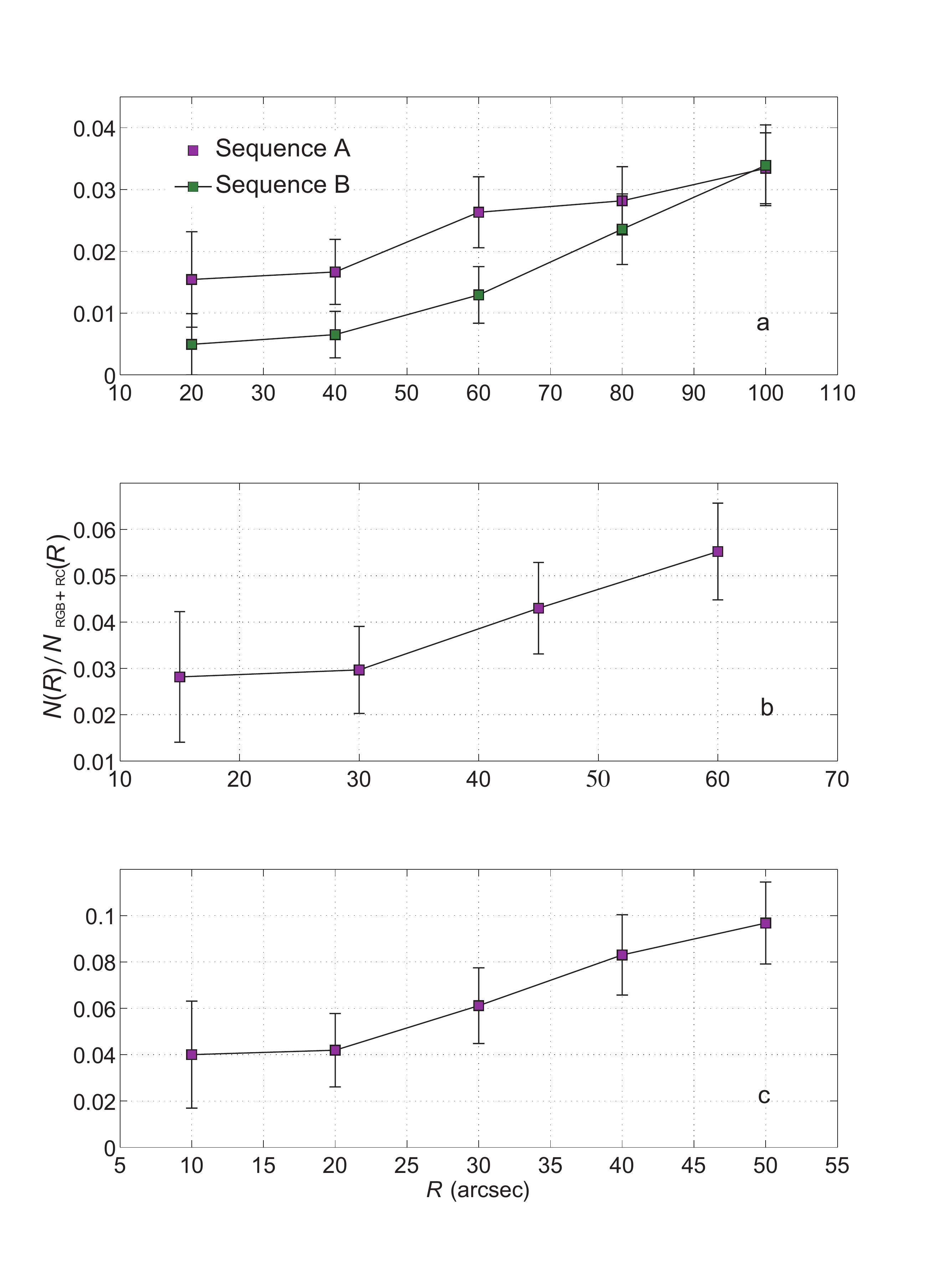}
\begin{center}
\caption{{\bf Normalized radial distributions of the young sequences with respect to normal cluster stars of similar luminosity. RGB: red-giant branch; RC: red clump.} a: NGC 1783; b: NGC 1696; c: NGC 411. The error bars reflect Poissonian 1$\sigma$ uncertainties.}
\end{center}
\label{F2}
\end{figure}

The more extended nature of the young populations suggests that they may have an external origin. Indeed, since the typical masses of NGC 1783, NGC 1696 and NGC 411 are only 1.8$\times$10$^5$, 5.0$\times$10$^4$ and 3.2$\times$10$^4$ solar masses\cite{Hunt03,McLa05}, respectively, they are insufficiently massive to efficiently capture the stellar winds from asymptotic-giant-branch stars\cite{Recc05}. Note that our NGC 1696 mass is based on extrapolation of the observed stellar luminosity function down to a stellar mass of 0.08 solar masses (the minimum mass for hydrogen fusion, somewhat depending on the star’s chemical composition), adopting a Kroupa-like initial mass distribution\cite{Krou01}. In addition, the observed upper limits in the mass-age diagrams populated by star clusters in the Magellanic Clouds are well-understood in terms of `size-of-sample' effects. They cannot be reconciled with initial cluster populations containing large numbers of young clusters with masses far greater than a million solar masses.

Adopting a Kroupa-like initial stellar mass distribution\cite{Krou01}, we estimate that the total stellar masses in the younger sequences, down to the hydrogen-burning limit, are 372 and 250 solar masses (NGC 1783 sequences A and B, respectively), 527 and 560 solar masses (NGC 1696 and NGC 411, respectively). Compared with the total cluster masses, the young sequences represent mere 0.2-2.0\% mass fractions. The age differences between the NGC 1783 stars in sequences A and B, and those on the main sequence are 440 million years and 520 million years, respectively. For NGC 1696 and NGC 411, the age differences between the clusters’ young and main sequences are 1.02 billion years and 1.06 billion years, respectively.

The Large and Small Magellanic Clouds, the host galaxies of our sample clusters, contain numerous, densely distributed giant molecular-gas clouds\cite{Fuku10,Hopk12}. Massive clusters like those targeted here quench their initial star-formation activity on timescales of a few tens of millions of years\cite{grijs07,grijs10} owing to the occurrence of Type II supernovae resulting from the deaths of the most massive first-generation stars\cite{Shus00,Conr11}. This leaves a cluster embedded in a gas-poor `cavity'. As the young star cluster moves through the interstellar medium, it could potentially accrete sufficient gas to fuel renewed star formation. In theory, secondary star formation can be triggered\cite{Conr11,Pfla09,Naim11} and proceed rapidly once the gas density reaches the relevant threshold, for sufficiently low temperatures. This would result in the appearance of a younger `simple stellar population'. However, to date the reality of this proposed idea has not been confirmed.

Although they are all contained within twice their host clusters’ core radii, the observed spatial distributions of all young stellar sequences in our sample clusters are more extended than those of their bulk stellar populations. This could indicate that these clusters may have accreted ambient gas, allowing star formation to proceed. We thus explored the expected gas-accretion rate as a function of the local gas density. Indeed, it appears possible for NGC 1783-like clusters to accrete enough gas to form new stars\cite{Conr11} (see Methods).

Almost all Galactic globular clusters host multiple stellar populations. However, it is still unclear whether these latter populations originate from young clusters that formed as single-age stellar populations. Our observations of secondary stellar populations in intermediate-age Magellanic Cloud clusters suggest that the same process giving rise to them may also explain the multiple stellar populations seen in at least some Galactic globular clusters. We aim at addressing this issue in a follow-up study.

Many star clusters in the Magellanic Clouds contain large numbers of stars occupying the parameter space at bluer colours and brighter luminosities than their main-sequence turn-off regions\cite{Milo09,Rich01}. These clusters include NGC 121, NGC 1652, NGC 1751, NGC 1795, NGC 1806, NGC 1846, NGC 1852, NGC 1917, NGC 1978, NGC 2121, NGC 2154, NGC 339, NGC 416, NGC 419, Hodge 7, Kron 3, Lindsay 1, and Lindsay 38. These brighter and bluer stars are usually dismissed as residual field-star contamination. However, if well-populated younger sequences are, in fact, embedded in this parameter space, the presence of such simple stellar populations indicates that many clusters may have experienced starburst events some time after their initial formation epoch. Among the clusters highlighted here, some -- including NGC 1806, NGC 1846, Lindsay 38 and NGC 419 -- appear to exhibit similar features as NGC 1783, NGC 1696 and NGC 411, although at a lower level of significance. Particularly for Lindsay 38 and NGC 419, the radial distributions of the bright stars beyond their main-sequence turn-off regions are known to be less concentrated than the bulk cluster stars with similar luminosities\cite{Glat08}, and hence a population of blue stragglers likely cannot explain the properties of all those stars. Some may have originated from gas accretion. These clusters will be targeted in our follow-up studies. Our discovery of clear, well-populated young sequences in NGC 1783, NGC 1696 and NGC 411 has revealed that star clusters may indeed have the capacity to accrete gas from their environment. This could, in fact, be the most important route to form secondary stellar populations in young massive clusters.

\begin{addendum}
 \item[Acknowledgements] We thank F. Ferraro, P. Kroupa and X. K. Liu for discussions. Partial financial support for this work was provided by the National Natural Science Foundation of China through grants 11073001, 11373010 and 11473037. C.L. was also partially supported by the Strategic Priority Research Programme `The Emergence of Cosmological Structures' of the Chinese Academy of Sciences (grant XDB09000000). A.M.G. is funded by a National Science Foundation Astronomy and Astrophysics Postdoctoral Fellowship under Award No. AST-1302765. C.-A.F.-G is supported by National Science Foundation grant AST-1412836.
  \item[Author Contributions] C.L., R.d.G. and L.D. jointly designed and coordinated this study. C.L. performed the data reduction. C.L., R.d.G., Y.X. and Y.H. collaborated on the detailed analysis. L.D. provided ideas that improved the study’s robustness. A.M.G. and C.-A.F.-G led the theoretical analysis of the gas-accretion physics. All authors read, commented on and jointly approved submission of this article.
 \item[Author Information] Reprints and permissions information is available at www.nature.com/reprints. Correspondence and requests for materials should be addressed to C.L. (joshuali@pku.edu.cn or licy@pmo.ac.cn).
 \item[Competing Interests] The authors declare that they have no
   competing financial interests.
\end{addendum}

\newpage
\section*{\large Methods}

\section*{Observational Data}

NGC 1783: The NGC 1783 observations were obtained as part of Hubble Space Telescope (HST) general-observer programme GO-10595 (principal investigator: P. Goudfrooij), using the Advanced Camera for Surveys/Wide Field Channel (ACS/WFC). The cluster was observed through the F435W and F814W filters (with central wavelengths of 435 nm and 814 nm, respectively), which correspond approximately to the Johnson-Cousins $B$ and $I$ bands, respectively, and which will be referred to as such henceforth. Short-exposure images were observed for 90 s in the $B$ band and 8 s in the $I$	 band, while long-exposure images were observed for 680 s in both bands.

Because NGC 1783 has an extended core (see below), the cluster region occupies almost the entire image. Therefore, we obtained an additional set of observations towards the southeast of NGC 1783 as representative field region (HST programme GO-12257; principal investigator: L. Girardi). Its centre is located at a distance of more than 300 arcseconds from the cluster centre, so that it is unlikely significantly affected by tidally stripped cluster stars. The data sets pertaining to the field were also obtained with the ACS/WFC in the F435W and F814W filters. Their total exposure times are 700 s and 720 s for the $B$ and $I$ bands, respectively. These exposure times are sufficient to resolve the young sequences.

NGC 1696: The NGC 1696 data sets were also obtained as part of HST programme GO-10595, using the ACS/WFC. The cluster was observed through the F435W and F814W filters. The images of NGC 1696 are also composed of short- and long-exposure frames. The long (short) exposure times in the $B$ and $I$ bands were 680 (90) s and 680 (8) s, respectively. Since NGC 1696 is not as extended as NGC 1783 (see below), a representative field region was selected close to the edge of the NGC 1696 images. Few studies to date have targeted NGC 1696. Therefore, we derived the physical parameters of NGC 1696 ourselves.

NGC 411: The data sets of NGC 411 were obtained as part of HST programme GO-12257, using the Wide Field Camera-3 (WFC3). The cluster was observed through the F475W and F814W filters. The F475W filter is centered at a wavelength of 475 nm; its transmission curve also corresponds approximately to that of the Johnson B band. The exposure times in both the $B$ and $I$ bands were 700 s. NGC 411 has a relatively small core (see below), which allowed us to select a representative field region close to edge of our science images.

\section*{Photometry and Data Reduction}

We used two independent software packages to perform point-spread-function (PSF) photometry, including {\tt DAOPHOT}\cite{Davi94} within the IRAF environment and {\tt DOLPHOT}\cite{Dolp13,Dolp11a,Dolp11b}. Our stellar catalogues are based on the {\tt DAOPHOT} results. We performed the {\tt DOLPHOT} analysis for comparison, to ensure that our final photometry is not biased.

The {\tt DAOPHOT}-generated raw stellar catalogue contains a sharpness parameter, which describes the goodness of the PSF fit\cite{Davi94}. For a `good' star, the sharpness should be close to 0. We thus constrained our sample to stars with a sharpness between -0.5 and 0.5, which removed approximately 4\% of objects from our catalogue. We carefully checked the resulting colour-magnitude diagram and found that the main features of interest were not affected by this selection. For NGC 1783 and NGC 1696, we merged the stellar samples resulting from the short and long exposure times, carefully cross-referencing both catalogues to avoid duplication of objects in the combined output catalogue. For NGC 1783, the short-exposure-time catalogue contributes very little to sequences A and B. For NGC 1696, the short-exposure-time stars contribute only marginally to the feature of interest.

\section*{Determination of the Cluster and Field Regions}

We divided the stellar spatial distribution into 15-20 bins along both the right ascension ($\alpha_{\rm J2000}$) and declination ($\delta_{\rm J2000}$) axes. We varied the bin numbers to ensure that statistical scatter would not significantly affect the shape of the number-density distributions. We used a Gaussian function to fit the latter along both axes and defined the closest positional coincidence of both Gaussian peaks as the cluster centre. The NGC 1783 cluster centre is located at $\alpha_{\rm J2000} = 04^{\rm h}59^{\rm m}08.47^{\rm s}$, $\delta_{\rm J2000} = -65^{\circ}59'17.81''$. For NGC 1696, the centre coordinates are $\alpha_{\rm J2000} = 05^{\rm h}02^{\rm m}11.16^{\rm s}$, $\delta_{\rm J2000} = -67^{\circ}59'07.66''$, while for NGC 411, $\alpha_{\rm J2000} = 01^{\rm h}07^{\rm m}56.22^{\rm s}$, $\delta_{\rm J2000} = -71^{\circ}46'04.40''$. Our NGC 1783 and NGC 411 cluster centres are very close to the cluster centres determined previously\cite{Werc11,Rafe05}. We used a Monte Carlo method to examine the spatial distribution of the NGC 1696 stars and estimate the areas of annuli at different radii. The stellar number density in each annulus is $N(R)/A(R)$, where $N(R)$ is the number of stars observed in an annulus with radius $R$, and $A(R)$ is the area of the annulus. We defined the clusters' (2D-projected) core radii as those radii where the density profiles drop to half the respective central densities. We selected the areas contained within 2 core radii as the cluster regions. NGC 1783 has a large core radius (45-50 arcsec, which is identical to that adopted by ref. 35, although our core radius is slightly larger than their value of 36.7 arcsec. We compared our catalogue with theirs and found that our photometry is deeper and, hence, contains more faint stars). The core radius of NGC 1696 is smaller ($\sim$30 arcsec), while the core radius of NGC 411 is approximately half that of NGC 1783 (20-25 arcsec). Hence, the cluster radii we adopted for NGC 1783, NGC 1696 and NGC 411 were 100 arcsec, 60 arcsec and 50 arcsec, respectively. We selected these radii as cluster radii, because of the need to avoid background contamination as much as possible, while simultaneously ensuring statistically robust results. In Extended Data Figure 1, we present the stellar number-density profiles.

\setcounter{figure}{0}

\begin{figure}
\hspace{-0.4cm}\centering
\includegraphics[width=1.0\columnwidth]{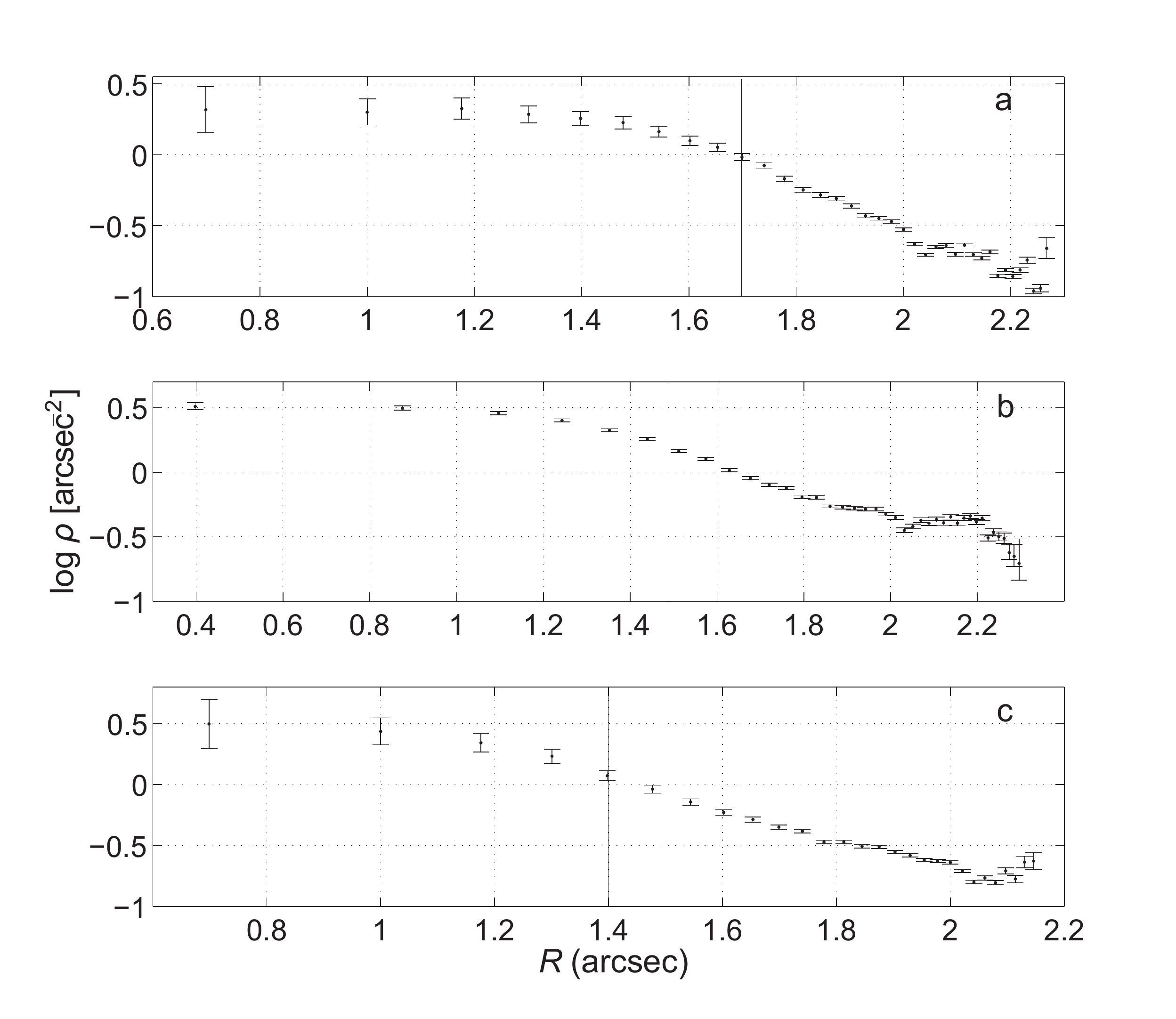}
\begin{center}
\caption{{\bf Normalized radial distributions of the young sequences with respect to normal cluster stars of similar luminosity. RGB: red-giant branch; RC: red clump.} a: NGC 1783; b: NGC 1696; c: NGC 411. The error bars reflect Poissonian 1$\sigma$ uncertainties.}
\end{center}
\label{ED1}
\end{figure}

For NGC 1783, we synthesized the field region’s colour-magnitude diagram based on a combination of our observations of the cluster’s periphery on the image containing the cluster (`field 1') and the field region towards the southeast (`field 2'). Because the features of interest are very bright, we only consider stars with B $\le$ 23 mag. For these stars, NGC 1783 is compact (its core radius is 25 arcsec) and field 1 indeed adequately represents the background. We found that the eastern part of field 2 exhibits a clear, slightly brighter stellar sequence parallel to the main sequence, indicating a significant population of unresolved binary systems. Since the field’s stellar number density is low, blending of unrelated stars along the line of sight is negligible; instead, this binary population may reflect contamination by a nearby star-forming region. We hence selected the eastern part of field 2 as representative field region. We selected three rectangles, each covering an area of 1000$\times$1000 pixels$^2$ ($\sim$50$\times$50 arcsec$^2$), near the edge of field 1, as well as four rectangles from the western part of field 2 (covering the same area), to construct a complete, combined field region. To assign equal weights to the stellar catalogues from both regions, we randomly selected ¾ of the full sample of stars detected in the four rectangles of field 2’s stellar catalogue. The combined stellar catalogue based on these seven rectangles represents the synthesized field region’s colour-magnitude diagram. The cluster region is roughly 1.6 times larger than the field region (see the left-hand panel of Extended Data Figure 2).

For NGC 1696, we selected an area of 600$\times$4000 pixels$^2$ ($\sim$30$\times$200 arcsec$^2$) near the edge of the image as representative field region. The NGC 1696 cluster region is roughly 1.7 times larger than the field region. The selected cluster and field regions pertaining to NGC 1696 are shown in the middle panel of Extended Data Figure 2.

For NGC 411, we selected a field region covering an area of 800$\times$3500 pixels$^2$ ($\sim$32$\times$140 arcsec$^2$) from the cluster’s periphery. We avoided regions that were located close to the edge of the image, where the photometric quality is markedly inferior, likely owing to the relatively large offsets between the individual science images, combined with instrumental propagation effects. Although our selection may include some tidally stripped cluster stars, this does not affect the magnitude range of interest, which is bright. The NGC 411 cluster region is roughly 1.8 times larger than the field region. The selected cluster and field regions of NGC 411 are shown in the right-hand panel of Extended Data Figure 2.

\begin{figure*}
\hspace{-0.4cm}\centering
\includegraphics[width=2.0\columnwidth]{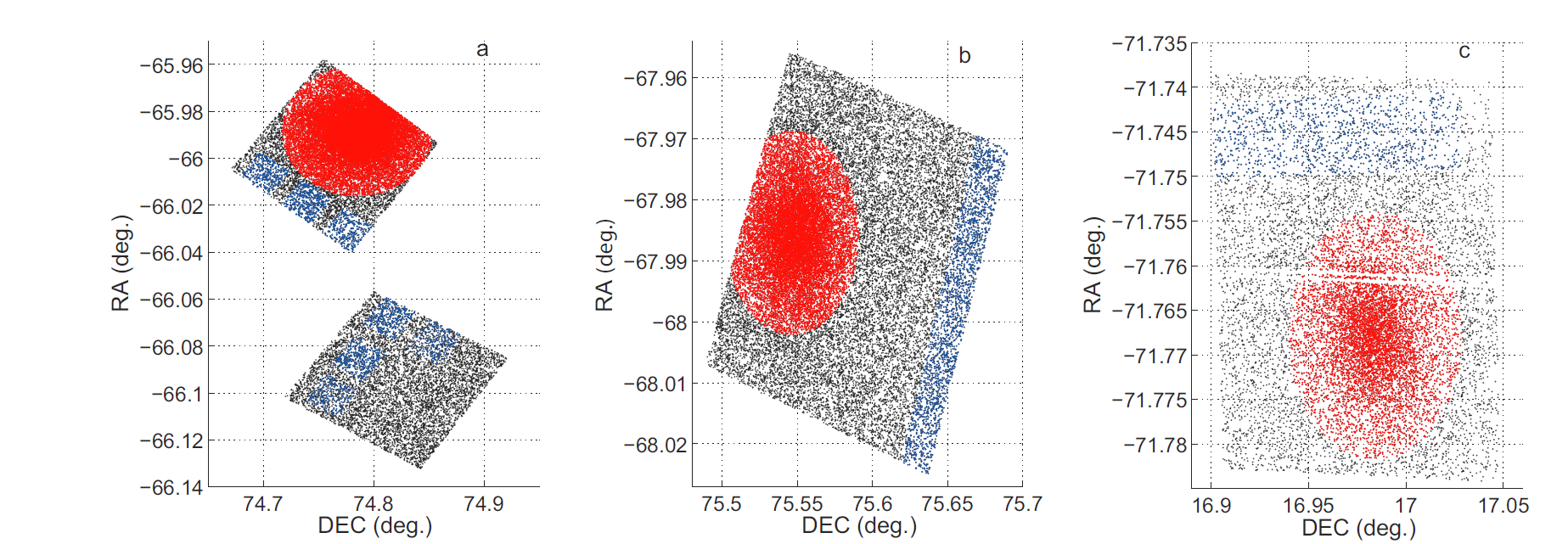}
\begin{center}
\caption{{\bf Spatial distributions of cluster and field stars. Red and blue points represent cluster stars and the adopted field stars, respectively.} a. NGC 1783; b. NGC 1696; c. NGC 411.}
\end{center}
\label{ED2}
\end{figure*}

\begin{figure*}
\hspace{-0.4cm}\centering
\includegraphics[width=2.0\columnwidth]{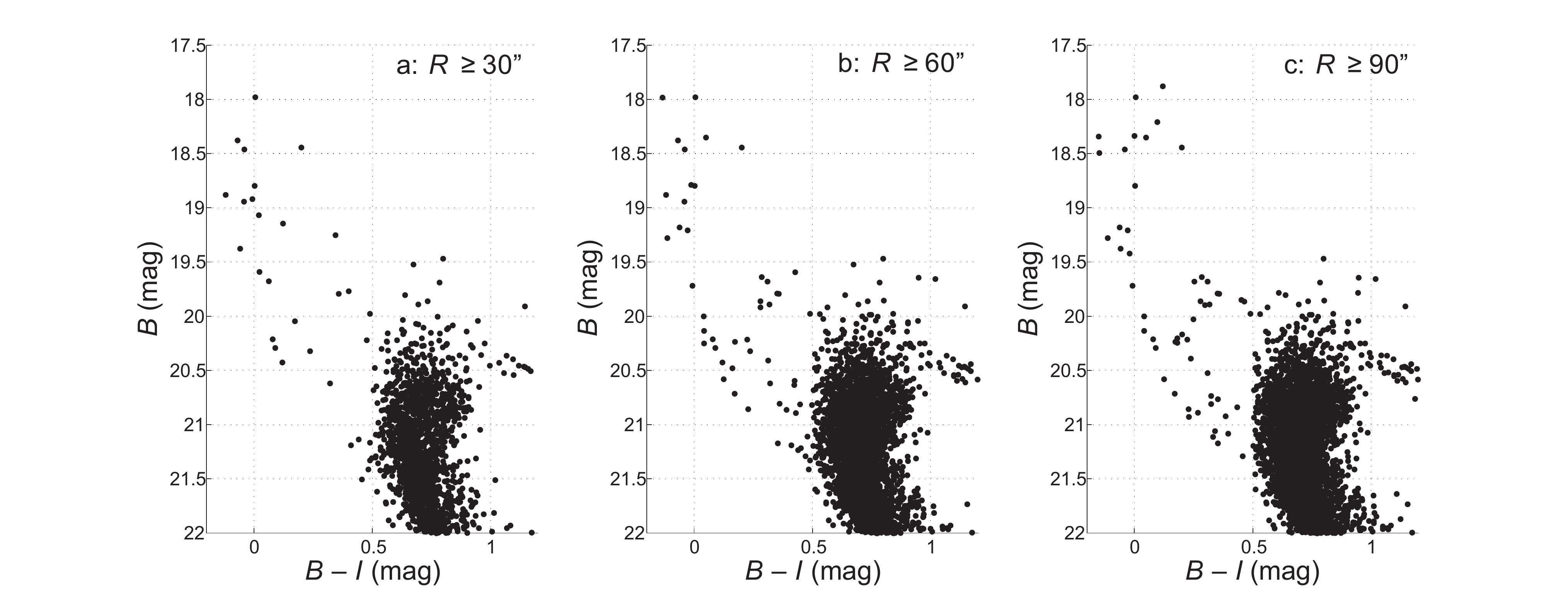}
\begin{center}
\caption{{\bf Field-star decontaminated colour-magnitude diagrams for samples of stars at different radii in NGC 1783.}  a. $R\geq30''$; b. $R\geq60''$; c. $R\geq90''$.}
\end{center}
\label{ED3}
\end{figure*}

\section{Reducing Background Contamination and Isochrone Fitting}

Once we had obtained the cluster and background colour-magnitude diagrams, we generated a common magnitude$\times$colour grid with cell sizes of 0.5$\times$0.25 mag$^2$, spanning the ranges from ($B - I$) = -2.5 mag to 3.5 mag and from $B$ = 16 mag to 27 mag. This range is sufficiently large to cover the full colour-magnitude diagrams of our target clusters. The cell size is relatively large for the main sequences, because it was specifically designed to be practically useful in the regions occupied by the young, blue sequences, where the stellar number density is lower. We counted the number of background stars in each cell and calculated the number of possible contaminating stars in the same cell (corrected for differences in areas covered), which we then randomly removed. We confirmed that varying the cell sizes from 0.3$\times$0.15 mag$^2$ to 0.5$\times$0.25 mag$^2$ for NGC 1783 would not affect the significance of any of the features of interest. For NGC 1696, the typical practically useful cell sizes range from 0.4$\times$0.4 mag$^2$ to 0.5$\times$0.5 mag$^2$, while for NGC 411, viable cell sizes range from roughly 0.3$\times$0.15 mag$^2$ to 0.5$\times$0.25 mag$^2$. Adopting much larger or smaller cell sizes would erase the observed sequences. We carefully examined the performance of our decontamination method and found that the observed features do not depend on the cluster or field regions selected: see Extended Data Figures 3 and 4, where we use NGC 1783 as benchmark. In Extended Data Figure 3, we show the field-decontaminated colour-magnitude diagrams pertaining to three different samples of cluster members, at radii $R\geq30''$, $R\geq60''$ and $R\geq90''$. They all exhibit distinct younger sequences. In Extended Data Figure 4, we select representative field regions from different images (two from a separate image and one from the image which also contains the cluster itself). The observed young sequences remain clearly visible for all three background regions adopted. In Extended Data Figure 5, we show the decontaminated colour-magnitude diagrams resulting from adoption of different grid sizes. This figure shows that the observed features are almost independent of grid size. Extensive tests also showed that the sequences found in the NGC 1696 and NGC 419 colour-magnitude diagrams are similarly well-defined. This confirms that the observed sequences are physically real rather than caused by statistical sampling effects. We also found that, for the adopted cell sizes, the reduced colour-magnitude distributions are similar in appearance to the real background colour-magnitude diagrams. Our adopted method therefore performs adequately. From Extended Data Figure 6 (panels a, e and i), one can deduce that these bright sequences are indeed already embedded in the unreduced colour-magnitude diagrams.

\begin{figure}
\hspace{-0.4cm}\centering
\includegraphics[width=1.0\columnwidth]{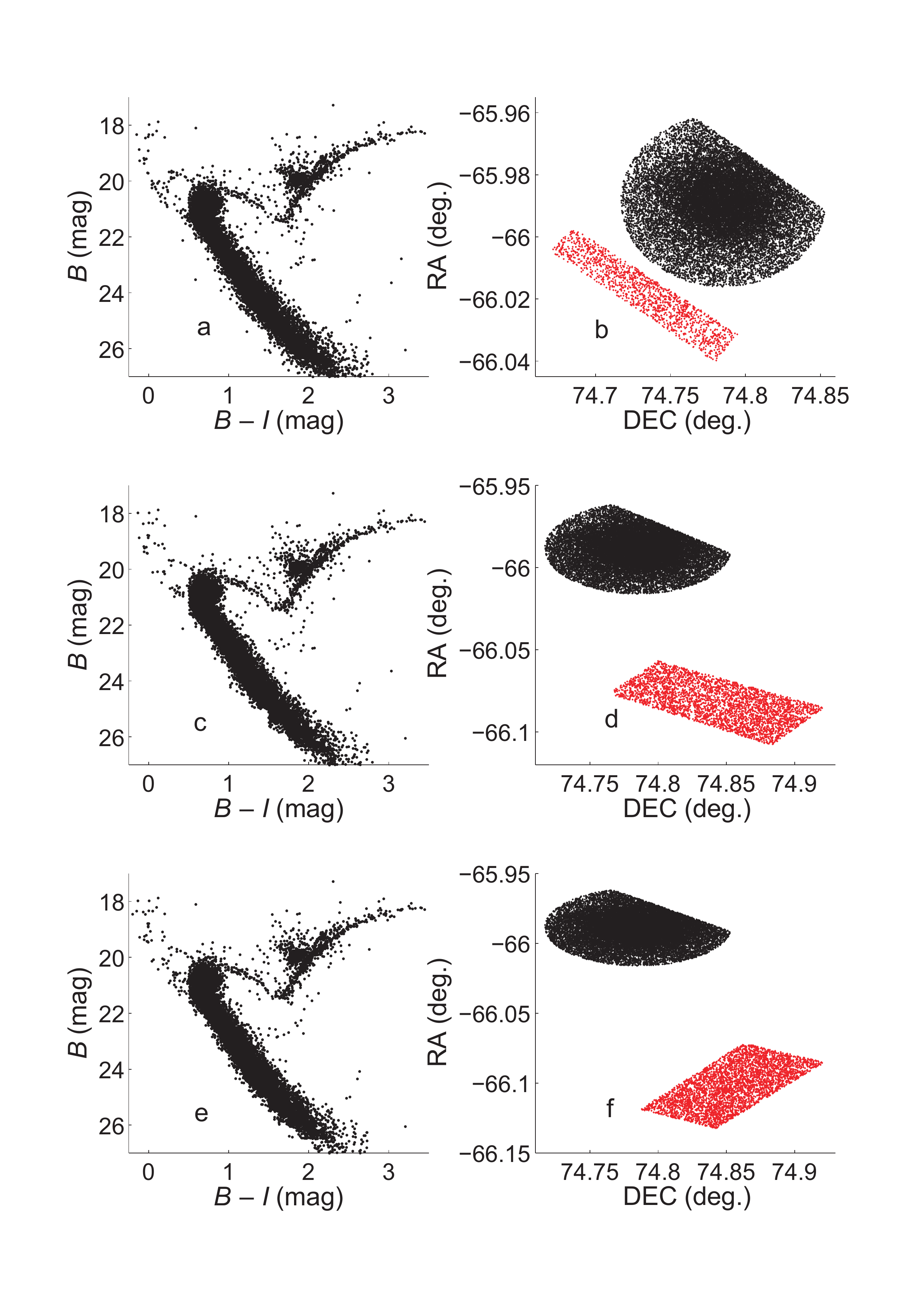}
\begin{center}
\caption{{\bf Field-star decontaminated colour-magnitude diagrams of NGC 1783 for three different adopted reference fields.} a, b. Resulting colour-magnitude diagram (a) based on field-star sample drawn from the image containing the cluster (b). c, d. As panels a and b, but for a representative field region taken from a separate image. e, f. As panels c and d, but for a different field region taken from the same, separate image.}
\end{center}
\label{ED4}
\end{figure}

\begin{figure*}
\hspace{-0.4cm}\centering
\includegraphics[width=2.0\columnwidth]{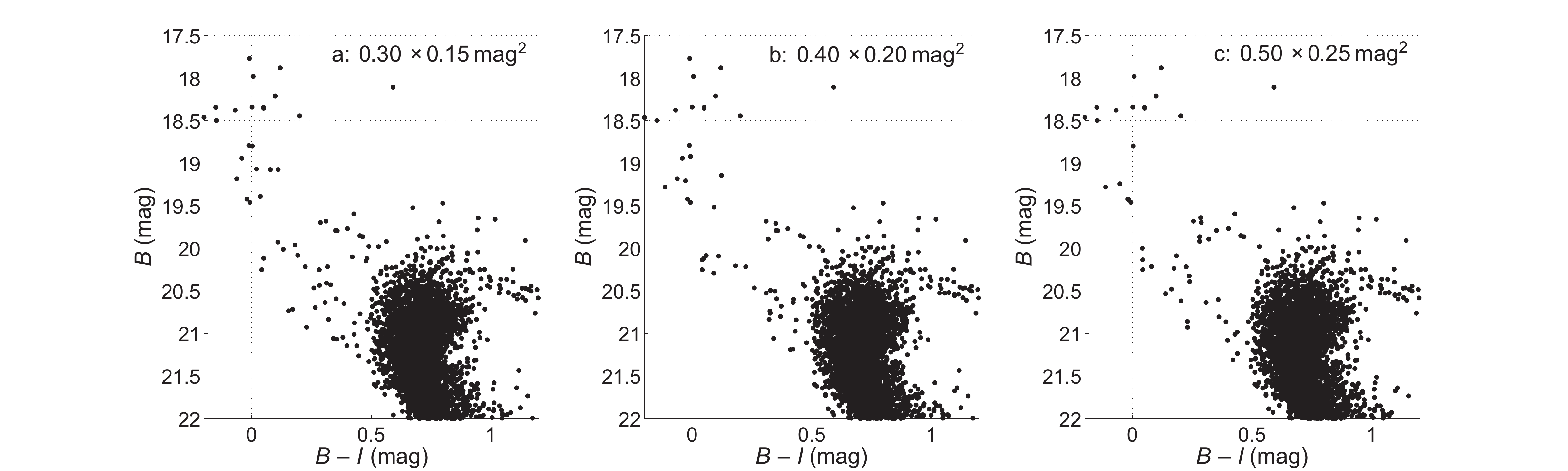}
\begin{center}
\caption{{\bf Field-star decontaminated colour-magnitude diagrams of NGC 1783 for three different grid sizes.} a. 0.30$\times$0.15 mag$^2$; b. 0.40$\times$0.20 mag$^2$; c. 0.50$\times$0.25 mag$^2$.}
\end{center}
\label{ED5}
\end{figure*}

\begin{figure*}
\hspace{-0.4cm}\centering
\includegraphics[width=2.0\columnwidth]{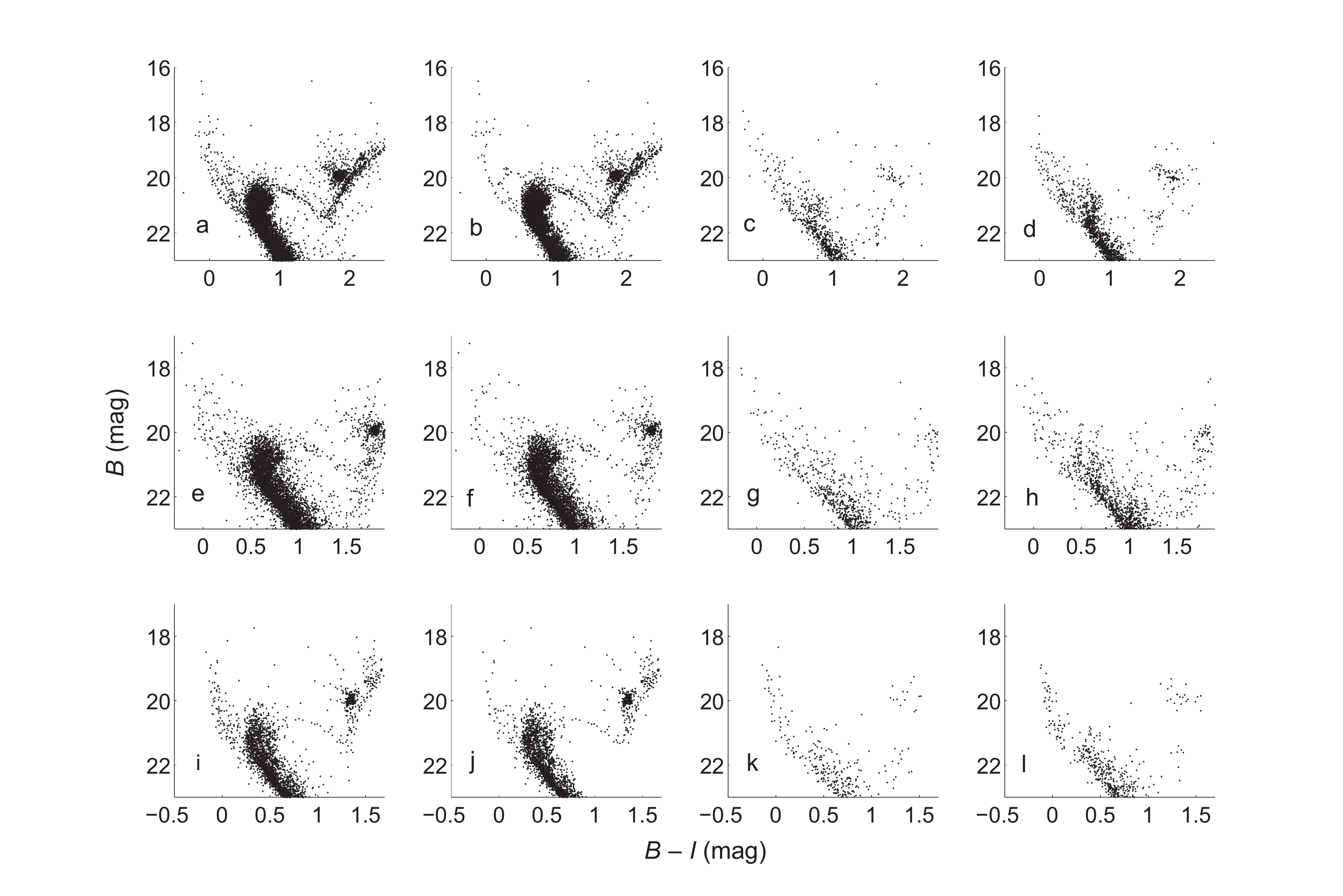}
\begin{center}
\caption{{\bf Colour-magnitude analysis.} First column: raw colour-magnitude diagrams. Second column: field-decontaminated colour-magnitude diagrams (also shown in Figure 1). Third column: colour-magnitude diagrams of the representative field regions. Fourth column: stellar colour-magnitude distributions that were removed from the raw catalogues. First row (a-d): NGC 1783; second row (e-h): NGC 1696; third row (i-l): NGC411. }
\end{center}
\label{ED6}
\end{figure*}

We obtained best fits to all observed sequences, including the clusters’ main sequences, based on matching the observations with theoretical stellar isochrones\cite{Mari08}. Similarly to other intermediate-age star clusters, NGC 1783, NGC 1696 and NGC 411 display extended main-sequence turn-off regions\cite{Milo09,Gira13}, which renders determination of their main-sequence ages difficult. However, it has been reported that ages of such intermediate-age star clusters can be constrained by consideration of their tight subgiant branches\cite{Li14,Bast15b} (but see ref. 39 for an opposing view), which represent the stellar evolutionary stage stars enter once they have exhausted the hydrogen in their cores through nuclear fusion. Indeed, all of our sample clusters exhibit tight subgiant branches. The final age determination yields $\log (t \mbox{ yr}^{-1})$ = 9.15, 9.17 and 9.14 for NGC 1783, NGC 1696 and NGC 411, respectively.

We next determined the ages of each of the blue sequences. For NGC 1783, sequences A and B are characterized by ages of $\log (t \mbox{ yr}^{-1})$ = 8.65 and 8.95, respectively. The NGC 1783 main-sequence stars, as well as those in sequences A and B, share the same metal abundance, $Z$ = 0.008 (40\% of solar metallicity), and visual extinction, $A_V$ = 0.06 mag. We adopted a true distance modulus for NGC 1783 of $(m - M)_0$ = 18.46 mag (corresponding to a distance of 49.2 kiloparsecs). The young sequence in NGC 1696 is adequately characterized by a $\log (t \mbox{ yr}^{-1})$ = 8.70 isochrone with the same metallicity as the cluster’s main sequence, $Z$ = 0.004 (20\% solar). However, the young sequence appears 0.06 mag bluer than the zero-age main sequence of the cluster’s bulk stellar population. An enhanced helium abundance ($Y$ = 0.330) can explain the colour offset\cite{Dott08}, where $Y$ = 0.256 for the cluster’s main sequence. The adopted extinction and distance modulus for NGC 1696 are $A_V$ = 0.10 mag and $(m - M)_0$ = 18.50 mag (50.1 kiloparsecs), respectively. The young sequence in NGC 411 has an age of $\log (t \mbox{ yr}^{-1})$ = 8.50. It has the same metallicity as the cluster’s main sequence, $Z$ = 0.002. Its young sequence is also very blue, which can again only be explained if it is characterized by an enhanced helium abundance of $Y$ = 0.400 compared with $Y$ = 0.252 for the cluster’s main sequence. Foreground extinction of $A_V$ = 0.25 mag is appropriate and our adopted true distance modulus is $(m - M)_0$ = 18.90 mag (60.3 kiloparsecs).

\section{Blue Straggler Stars as possible origin of the younger sequences?}

One possible explanation for these bright sequences is that they are composed of blue straggler stars. Therefore, we investigated their relative radial concentration with respect to stars that have similar luminosities. In NGC 1783, NGC 1696 and NGC 411, the latter stars are mostly red-giant-branch and red-clump stars: see Extended Data Figure 7.

\begin{figure*}
\hspace{-0.4cm}\centering
\includegraphics[width=2.0\columnwidth]{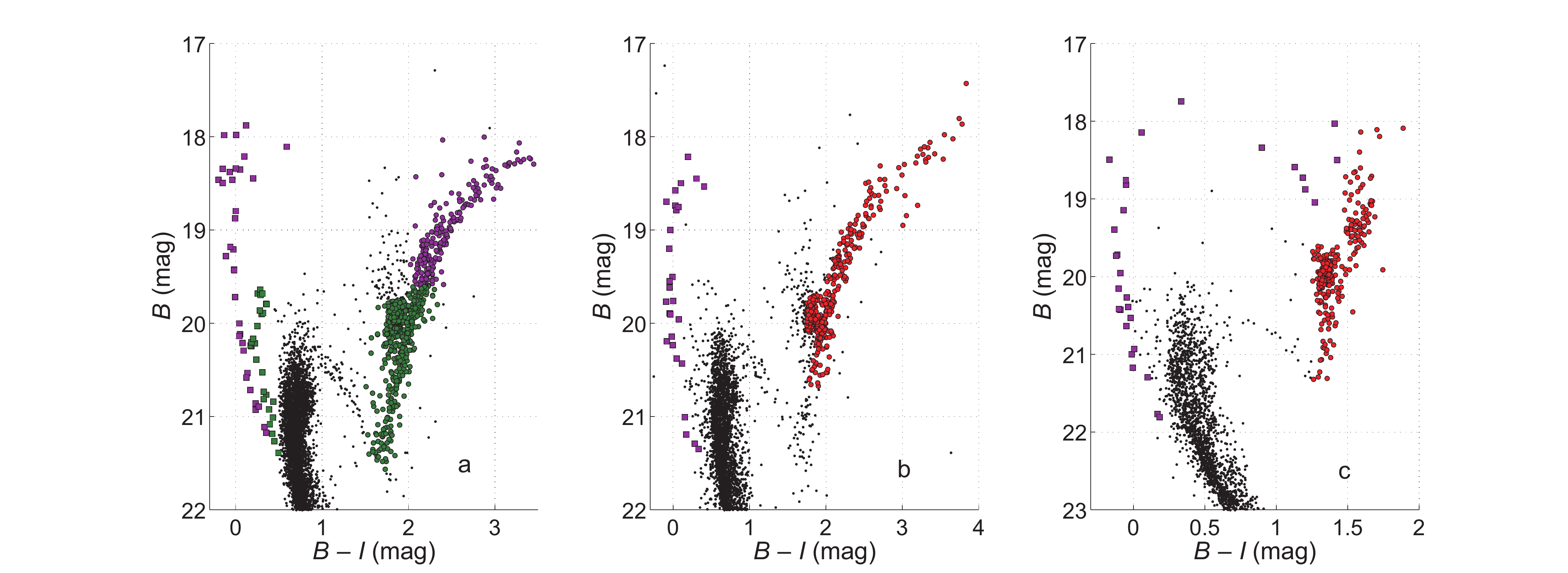}
\begin{center}
\caption{{\bf Colour-magnitude diagrams highlighting specific features.} a, Purple and dark green squares: stars in NGC 1783 sequences A and B, respectively. Dark green circles: corresponding red-giant-branch and red-clump stars, used for comparison with sequence B. The combination of dark green and purple circles represents the sample used for comparison with sequence A.
b. Purple squares: NGC 1696 young-sequence stars; red circles: corresponding red-giant-branch and red-clump stars used for comparison.
c. As the middle panel, but for NGC 411.
}
\end{center}
\label{ED7}
\end{figure*}

However, as shown in Figure 2, the stars defining the bright sequences are all less centrally concentrated than red-giant stars with similar luminosities. If they are genuine blue stragglers, irrespective of their origin, they are expected to be {\it more} centrally concentrated than similar-luminosity red-giant stars, because blue stragglers are expected to be more massive than red giants. Dynamical interactions are also unlikely to have redistributed all blue stragglers to the outskirts of our sample clusters, since the typical dynamical timescales\cite{BT98} are much longer than the clusters’ current ages. In addition, the flat cluster cores observed for NGC 1783, NGC 1696 and NGC 411 render the probability of core-collapse events having occurred very unlikely. Core collapse would produce a cuspy radial density profile\cite{Ashm98} and such a process is a prerequisite for the presence of a coeval collisional blue straggler population. The more extended radial distributions of the stars in the younger sequences compared with the dominant cluster populations argue against stellar collisions in the cluster cores having played an important role.

\section{Gas Accretion from the Interstellar Medium}

We followed the method of ref. 23 to estimate the regions of parameter space where an NGC 1783-like star cluster with a mass of 1.8$\times$10$^5$ solar masses and a half-mass radius of 11.4 parsecs\cite{Goud14} could accrete the required mass to form two additional generations of stars: one containing 250 solar masses in stars some 520 million years after the initial star-formation event, and a subsequent generation composed of 370 solar masses of stars 440 million years later. The key equations in ref. 23 are equations (3), (5) and (8). For all calculations we assumed a star-formation efficiency of 10\% (which implies that only 10\% of the available gas is converted into stars). We explored two processes through which a cluster can accrete gas from the interstellar medium\cite{Conr11}, one owing to gravity (`Bondi accretion'), and one due to sweeping up material as the cluster orbits around its host galaxy’s centre. The latter process involves some initial intracluster gas which interacts collisionally with the interstellar gas. The accretion rates from both of these processes depend on both the gas density, $n$, and the relative velocity of cluster and gas, $V$. Ram-pressure stripping can limit the accretion, and this process also depends on $n$ and $V$.

Extended Data Figure 8 shows the allowed (shaded) regions of parameter space in ($n$, $V$) that would enable a cluster to accrete the desired amount of gas in the period between star-formation events. This figure assumes accretion from a volume-filling interstellar medium over the entire duration between respective generations (Bondi accretion). The Bondi line is curved, because this accretion rate also depends on the gas sound speed, which we have assumed to be 10 km s$^{-1}$ (i.e., gas at a temperature of $\sim$10$^4$ K). Anything to the upper right of the `Ram' line will strip the gas from the cluster. Below that line, and for a given density, the Bondi line defines the upper limit to the cluster’s bulk velocity that would be allowed for the cluster to accrete this amount of gas. In other words at a velocity below the curve, the cluster would accrete gas more quickly, and, for instance, if the star-formation efficiency were less than 10\%, it could still form the same mass of stars over the same period. Conversely, at a given density, sweeping up of gas is more efficient at larger velocities, so the dashed line shows the lower limit to the velocity that would be required to reach the desired amount of gas.

\begin{figure}
\hspace{-0.4cm}\centering
\includegraphics[width=1.0\columnwidth]{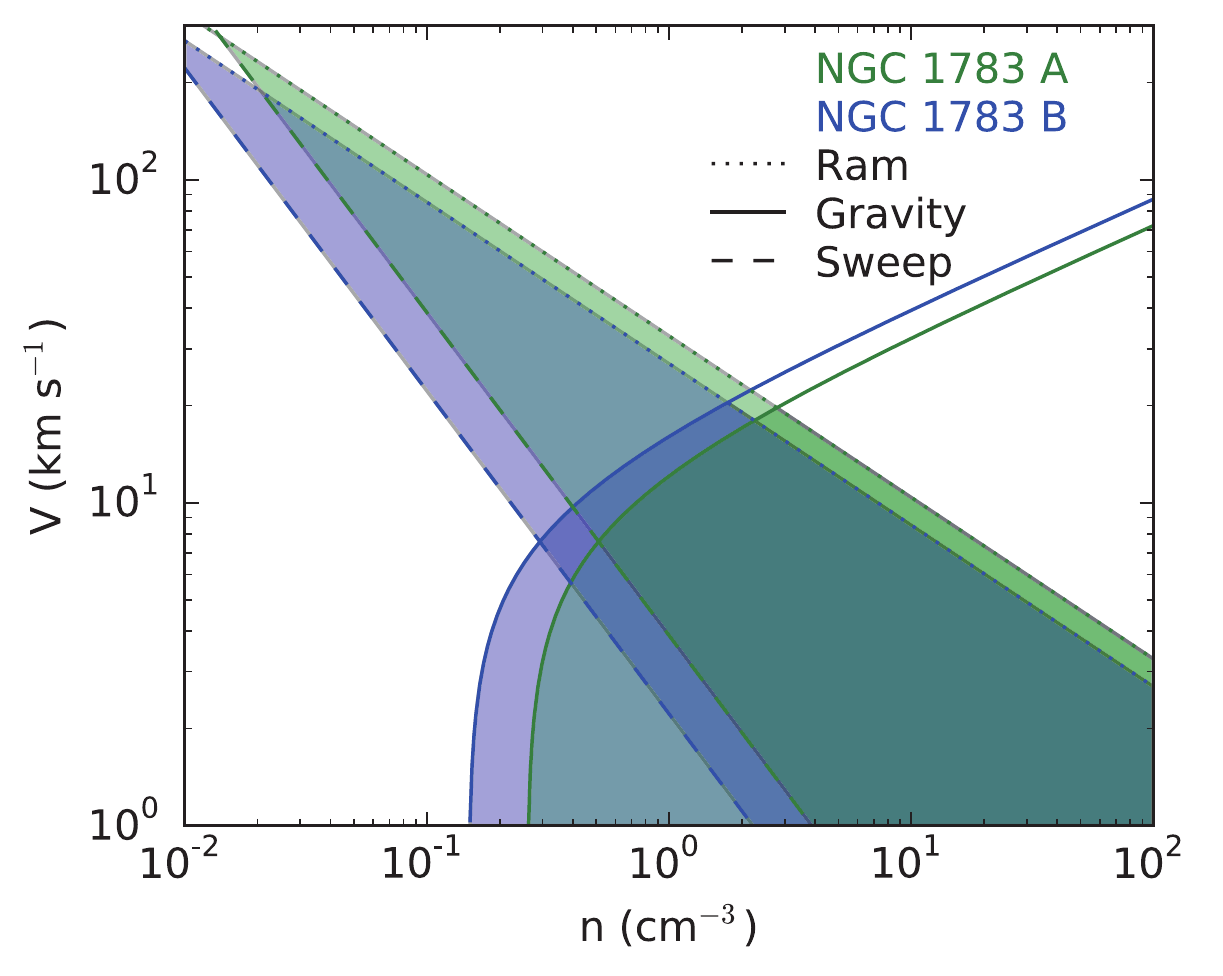}
\begin{center}
\caption{{\bf Gas-accretion diagnostic diagram.} The shaded regions indicate the parameter space where an NGC 1783-like cluster can accrete the required mass to form the two additional generations of stars, namely one of 250 solar masses over 520 million years (blue), and a second of 370 solar masses over 440 million years (green). We have assumed a star-formation efficiency of 10\% for all calculations in this figure. The regions to the right of the `Gravity' curves correspond to where Bondi accretion can accumulate at least the required mass and the regions to the right of the `Sweep' curves correspond to where accretion by collisional sweeping up of ambient interstellar gas by seed intracluster gas can accumulate at least the required mass. The parameter space above the `Ram' curves are excluded because ram pressure strips clusters of their gas in those regions.
}
\end{center}
\label{ED8}
\end{figure}

Extended Data Figure 9 shows two examples of how the accreted mass could accumulate over time for the Bondi regime (solid lines) and the sweeping regime (dashed lines). For the Bondi regime, we used a relative velocity of 4 km s$^{-1}$ and a density of 0.3 cm$^{-3}$. For the sweeping regime, we used a velocity of 50 km s$^{-1}$ and a density of 0.05 cm$^{-3}$ (which is similar to the rotation velocity at the position of NGC 1783 with respect to the Large Magellanic Cloud’s centre\cite{Mare02}). The data points show these masses, with 150 solar-mass error bars divided by the star-formation efficiency of 10\%.

\begin{figure}
\hspace{-0.4cm}\centering
\includegraphics[width=1.0\columnwidth]{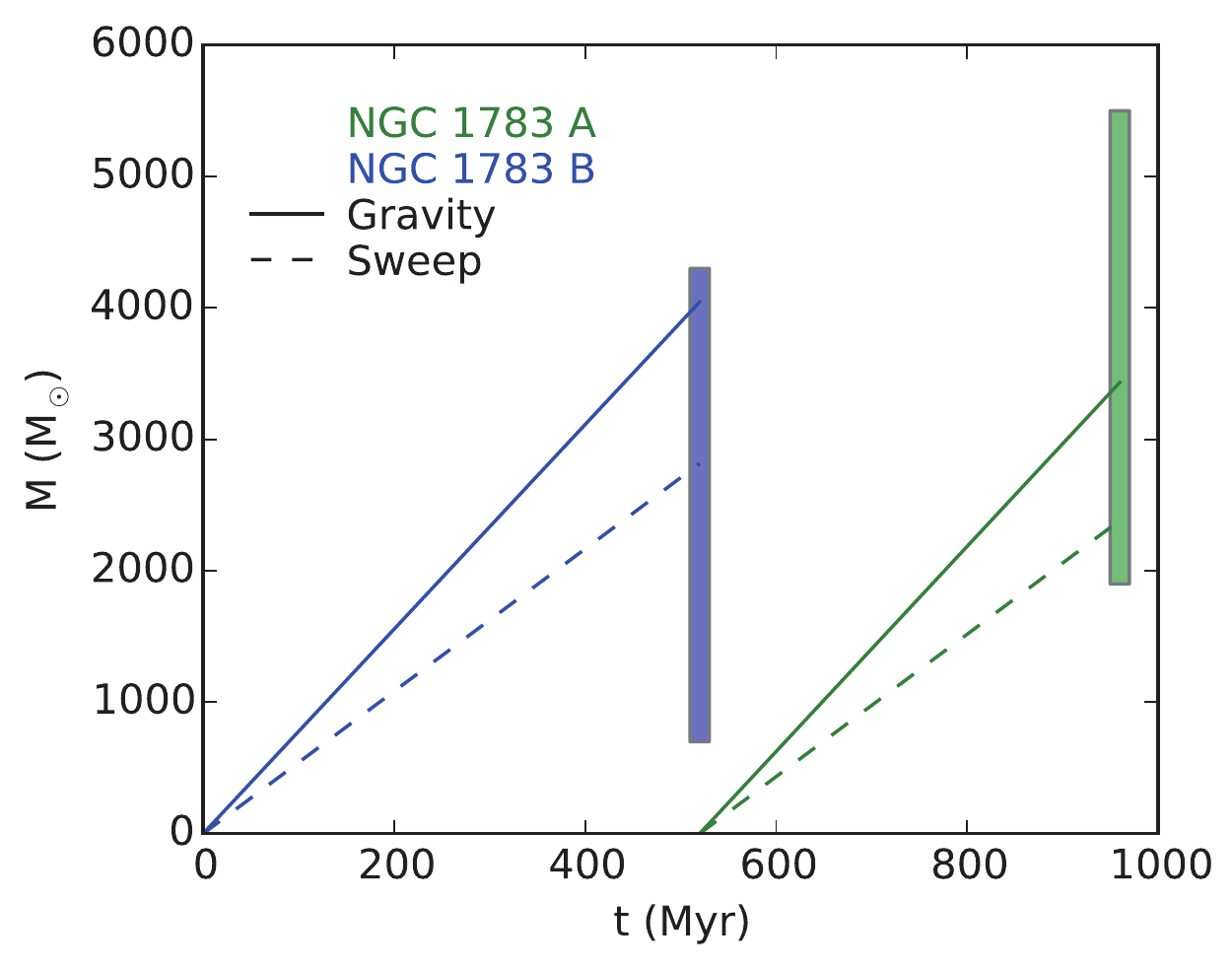}
\begin{center}
\caption{{\bf Gas mass accreted from the interstellar medium as a function of time.} We have adopted a star-formation efficiency of 10\% and calculated representative interstellar gas-accretion frameworks that can explain the stellar masses in the secondary sequences A and B in NGC 1783. For the Bondi regime (solid lines), we used a relative velocity of 4 km s$^{-1}$ and a density of 0.3 cm$^{-3}$. For the sweeping regime (dashed lines), we used a velocity of 50 km s$^{-1}$ and a density of 0.05 cm$^{-3}$. The blue and green data points indicate the stellar masses and age offsets of NGC 1783 sequences A and B, respectively. The error bars represent uncertainties in the star-formation efficiency of 10\%.-}
\end{center}
\label{ED9}
\end{figure}

\end{document}